\documentclass[aps,showpacs,showkeys,prl,reprint,superscriptaddress]{revtex4-1}
\usepackage{amsmath,nicefrac,units,multirow,dcolumn,float}
\usepackage[american]{babel}
\usepackage[T1]{fontenc}
\usepackage{natbib}
\usepackage[utf8]{inputenc}
\usepackage{graphicx}
\usepackage[authormarkup=none]{changes}

\usepackage{epsfig}
\usepackage{color}
\usepackage{srcltx}
\usepackage{subfigure}
\usepackage{dcolumn}
\usepackage{csquotes}
\usepackage{color}
\usepackage{srcltx}

\definecolor{gruen}{rgb}{0,0.4,0}

\begin{document}

\title{
Offset-corrected $\Delta$-Kohn-Sham scheme for the prediction of
X-ray photoelectron spectra of molecules and solids
}
\date{\today}

\author{Michael \surname{Walter}}
\email{Michael.Walter@fmf.uni-freiburg.de}
\affiliation{Freiburger Materialforschungszentrum, Universität Freiburg, Stefan-Meier-Straße 21, D-79104 Freiburg, Germany}
\affiliation{Fraunhofer IWM, MikroTribologie Centrum $\mu$TC, Wöhlerstrasse 11, D-79108 Freiburg, Germany}
\author{Michael \surname{Moseler}}
\affiliation{Freiburger Materialforschungszentrum, Universität Freiburg, Stefan-Meier-Straße 21, D-79104 Freiburg, Germany}
\affiliation{Fraunhofer IWM, MikroTribologie Centrum $\mu$TC, Wöhlerstrasse 11, D-79108 Freiburg, Germany}
\affiliation{Physikalisches Institut, Universität Freiburg, Herrmann-Herder-Straße 3, D-79104 Freiburg, Germany}
\author{Lars \surname{Pastewka}}
\affiliation{Fraunhofer IWM, MikroTribologie Centrum $\mu$TC, Wöhlerstrasse 11, D-79108 Freiburg, Germany}
\affiliation{Karlsruher Institut für Technologie, Institut für Angewandte Materialien, Engelbert-Arnold-Stra\ss e 4, D-76131 Karlsruhe, Germany}
  
\begin{abstract}
Absolute binding energies of core electrons in molecules and bulk materials 
can be efficiently calculated by spin paired density-function theory 
employing a $\Delta$ Kohn-Sham ($\Delta$KS) scheme corrected by offsets 
that are highly transferable. 
These offsets depend on core level and atomic species and can be 
determined by comparing $\Delta$KS energies to experimental molecular 
X-ray photoelectron spectra.
We demonstrate the correct prediction of absolute and 
relative binding energies on a wide range of molecules, metals and insulators.

\end{abstract}


\maketitle

\begin{figure}
  \includegraphics[width=\linewidth]{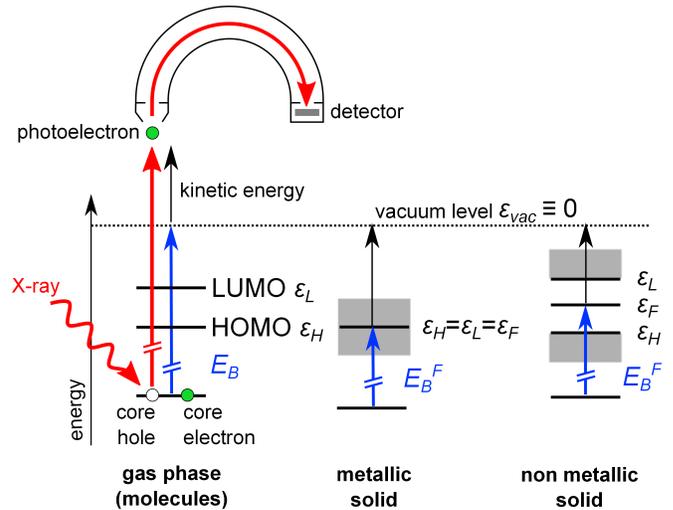}
  \caption{
    \label{fig:Energies}
    Schematic presentation of X-ray photoelectron spectroscopy and the relevant energy levels in core hole ionization.
    The core electron binding energy $\text{CEBE}=-E_B$ for molecules and $\text{CEBE}=-E_B^F$ for solids.
  }
\end{figure}

X-ray photoelectron spectroscopy (XPS) has become a valuable tool 
for the characterization of molecules and 
materials~\cite{Siegbahn:Book1969,Siegbahn:1973p493}. 
High energy photons ionize the system under study by 
removing an electron from a core level (see Fig.~\ref{fig:Energies}). 
The resulting energy loss can be interpreted as a direct measure 
of the core electron binding energy (CEBE). 
Core levels typically range between several tens ($\sim 80\text{eV}$) for
Al(2s)) to hundreds ($\sim 570\text{eV}$ for F(1s)) of eV 
and this large spread allows for a determination of the 
elemental composition of materials. 

Furthermore the exact CEBE of an atomic species can vary by several eV, 
depending on its local chemical environment. 
This ``chemical shift'' is an indirect measure for the configuration 
of the valence electrons that determine the atoms' bonds.
It forms the basis of the ESCA method 
(electron spectroscopy for chemical analysis)~\cite{Siegbahn:Book1969,Siegbahn:1973p493} that is routinely used for chemical material characterization.
In practice the assignment of experimental XPS lines 
relies on extensive tables of CEBEs measured in 
well-known compounds~\cite{Beamson:1992p295,Henke1993,Moulder1995,Crest:Book2000,NISTXPS}.
An \emph{ab-initio} prediction of core levels has 
turned out to be useful for the interpretation of experimental spectra 
to finer details~\cite{Bagus1965,Lindgren:2004p59}.
Such predictions are particularly fruitful for more complicated 
molecules or solids where atoms of the same elements are present 
in heterogeneous environments, for instance in minerals and glasses, 
where local coordination number as well as local bond lengths can 
vary from one atomic site to the other~\cite{Nesbitt:2014p271}.

In this letter, we provide an extensive benchmark of 
inexpensive density-functional theory (DFT) 
calculations~\cite{Martin:Book2004} within the 
$\Delta$-Kohn-Sham ($\Delta$KS) formalism~\cite{Triguero98prb}. 
We carry out $\Delta$KS calculations of {\em absolute} CEBEs for 
isolated systems and explore the potential of periodic $\Delta$KS 
to reproduce {\em absolute} CEBEs from solid state XPS experiments.
In view of future applications to large material systems 
we aim at the most efficient scheme and therefore our 
$\Delta$KS approach is based on spin-paired calculations. 
We demonstrate below, that in contrast to the general 
believe~\cite{Haerle01prb}, 
periodic DFT can be used for the prediction of
{\em absolute} CEBE values with good accuracy.
Errors that are associated with the spin-paired treatment 
and with inaccuracies in the exchange-correlation potential can
easily be compensated by offsets $\delta(X(nl))$ that 
depend merely on the element type $X$ and 
the core level $nl$ (e.g. $\delta(C(1s))$ is the offset of 
the carbon 1s core level). 
These highly-transferable offsets are virtually independent
of the chemical environment of the respective species $X$ and 
do not vanish for spin-polarized calculations.

{\em Theory.} 
Despite the common interpretion as a single particle energy, the CEBE 
\begin{equation}
  \label{eq:ech}
  -E_B = E_0 - E^+_{\rm ch}
\end{equation}
is
the difference between the ground state energy of a system $E_0$ and
the energy of an (metastable) excited state $E^+_{\rm ch}$ of the 
same system with one electron 
less~\cite{Egelhoff:1987p253,Walter08njp}.
The definition of the ionized core-hole state 
itself relies on an effective
single particle picture and is thus not unambigeous.
Nevertheless, the $\Delta$KS method,
which is defined through a 
partly filled Kohn-Sham orbital,
has been shown to correlate well with experiment
\cite{Triguero98prb,Takahata03jesrp,Takahashi04jpc}.

A molecular CEBE can be directly compared to experiment
as the energy scales of both the 
neutral ground-state and the ionized state 
that contains the core hole are well defined.
The reference energy of such finite systems is the vacuum level 
$\varepsilon_\text{vac}\equiv 0$, 
i.e. the energy of an electron at infinite distance from the molecule 
(see Fig.~\ref{fig:Energies}).
The vacuum level also serves as a convenient reference for
the energy needed to remove the most weakly bound valence electron
(the ionization potential, IP) and for the 
energy gained by an extra electron
(the electron affinity, EA).
These many particle energies unambiguously define the 
single particle energies of the highest occupied (HOMO) and 
the lowest unoccupied molecular orbital (LUMO), 
$\varepsilon_H=-\text{IP}$ and $\varepsilon_L=-\text{EA}$, 
respectively \cite{Baerends13pccp}.
Eq.~\eqref{eq:ech} can be directly applied when the electron 
is ejected into the vacuum 
surrounding the molecule.

The vacuum level is not an easily accessible reference for extended systems. 
There the CEBE is usually reported relative 
to the Fermi level $\varepsilon_F$ 
\cite{Riga77mp}, that is the energy threshold up to which the single 
particle levels are filled:
\begin{equation}
  \label{eq:echF}
  E^F_{B} = E^+_{\rm ch} - E_0 + \varepsilon_{F}
  \; .
\end{equation}
The  charged state energy $E^+_{\rm ch}$ in
Eq.~\eqref{eq:echF} poses a problem in simulations as
extended system typically have to be modeled 
using periodic boundary conditions.
These require a neutral unit cell to avoid
interaction of the charge with its periodic images.
This problem can be overcome by artificially neutralizing the system, 
by either adding an extra valence 
electron~\cite{Pehlke93prl,Snis99prb,Haerle01prb} or by 
adding a constant background charge~\cite{Haerle01prb,Titantah05carbon,Ljungberg11jesrp,Susi14bjn}.

Adding a valence electron changes the total energy of the system by the 
LUMO energy $\varepsilon_L$\footnote{
We regard the LUMO energy $\varepsilon_L$ as constant
under the addition of the extra electron, 
an approximation that becomes exact in the limit of infinite supercells.} 
such that the neutral system containing the core hole
has the energy
$E^0_{\rm ch} = E^+_{\rm ch} + \varepsilon_L$ and
the binding energy of the core hole for 
extended systems becomes
\begin{equation}
  \label{eq:echF0}
  E^F_{B} = E^0_{\rm ch} - E_0 + \varepsilon_{F} - \varepsilon_L
  \; .
\end{equation}

In metallic systems, many of the single particle energies coincide as
$\varepsilon_F=\varepsilon_H=\varepsilon_L < 0$
(see Fig.~\ref{fig:Energies}).
In this case, 
the core hole binding energy relative to the Fermi level simply 
becomes $E^F_{B} = E^0_{\rm ch} - E_0$.
This quantity is also straightforward to measure in experiments
according to Eq. (\ref{eq:echF}),
since for metallic systems the Fermi level is given by 
the electrostatic potential (voltage) of the metallic specimen. 

The situation complicates for non-metals.
For systems with a finite band gap $\varepsilon_L-\varepsilon_H$, 
$\varepsilon_F$ 
lies somewhere inside of the gap. 
The exact position depends on temperature, is 
strongly influenced by impurities~\cite{Sze:Book1981} and can be 
modified by the presence of surface layers~\cite{Zhong12pccp}.
Measured binding energies for non-metals therefore tend to 
vary by many eV from experiment to experiment.
Computationally, the situation is further complicated because the 
empty (Kohn-Sham) energy levels obtained from density functional theory 
calculations (based on the usual gradient corrections)
are typically too low in energy leading a significant underestimation of the band gap.

Chemical shifts are independent of these complications since the 
energy reference cancels out.
In application of a chosen single particle picture,
these shifts were analyzed by contributions that are attributed 
to differences 
between two atomic sites already present in the “initial state” or 
emergent in final state of the ionization process~\cite{Egelhoff:1987p253}.
%
%
In Kohn-Sham DFT, the method of choice for large molecules and 
solids due to its computational efficiency, 
approximate CEBEs are obtained directly from the 
eigenvalues of core states~\cite{
Koehler04prb,Gandubert08ct,Vogel12prb}.
This approach is especially popular in solid state applications 
since it circumvents the treatment of
charged ionized states in periodic supercells.
It was shown, however, that the neglect of relaxation 
in the final state 
results in appreciable errors including predictions of 
erroneous shift directions~\cite{Pehlke93prl}
and prohibits the determination of 
reliable absolute CEBEs~\cite{Haerle01prb}.  
Within the single particle picture, relaxation effects can 
be included via
Slater-Janak transition state theory, where the orbital
energy of a partly occupied core hole state
is interpreted as CEBE
~\cite{Chong:1995p486,Chong:1995p1842,Gandubert08ct}.

In order to fully capture the final state effects, eq. (\ref{eq:ech})
has to be applied. 
The simplest approach to obtain $E^+_{\rm ch}$ or $E^0_{\rm ch}$
is the $Z+1$ or equivalent-core 
approximation~\cite{Martensson:1979p791,Johansson:1980p4427}, 
where the excited nucleus is replaced by the next element in the 
periodic table. This approach is covered by ground state DFT,
but assumes the density change to be located exactly
at the position of the nucleus.
While stretching the DFT's validity \cite{Egelhoff:1987p253}, 
the most accurate results are obtained by modelling the core hole as 
a partly occupied atomic state either in all electron
\cite{Triguero98prb,Snis99prb,Takahata03jesrp,Titantah05carbon,Susi15prb}, 
effective core or pseudopotentials, 
\cite{Pehlke93prl,Haerle01prb,Takahashi04jpc}
or within the projector augmented wave method
and the frozen core approximation
\cite{Koehler04prb,Ljungberg11jesrp}.
These approaches lead to very similar results and
yield differences of around 
$50\,\text{meV}$~\cite{Koehler04prb}. 

\begin{figure*}[ht]
  \includegraphics[width=\linewidth]{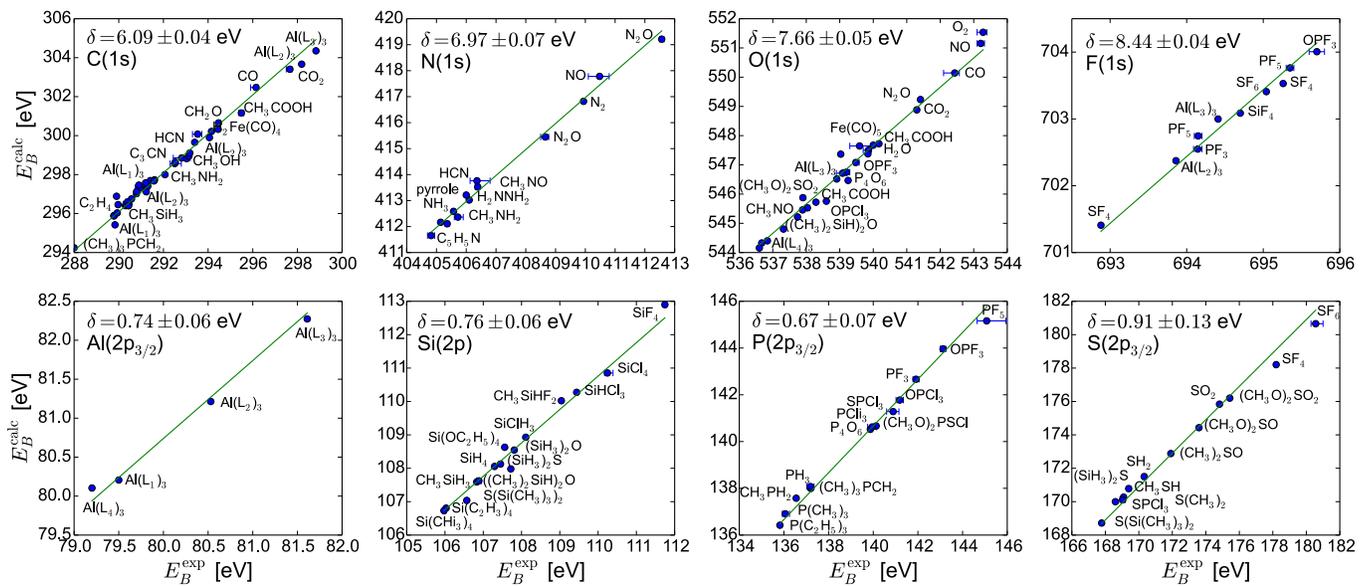}
  \caption{
    \label{fig:molecules}
    Core electron binding energies obtained with the PBE functional 
    compared to 
    experimental values for molecules in the gas-phase.
    The constant offset $\delta$ for each element corresponding to 
    the straight line is given (see text).
    The exact numbers are listed in Supplemental Material \cite{supplemental}.
}
\end{figure*}

{\em Methods.}
The following DFT calculations were carried out with the GPAW
\cite{Mortensen05prb,Enkovaara10jpc} package,  an implementation of
the projector augmented wave (PAW) 
method~\cite{Bloechl:PAW}.
We use the
configuration space grid implementation and apply a grid
spacing of \unit[0.2]{\text{\AA}} to represent the smooth wave functions 
unless noted otherwise. 
The exchange correlation energy was approximated by the
generalized gradient correction by Perdew et al. (PBE)~\cite{Perdew96prl}
and additional calculations were carried out
in the local density approximation (LDA) \cite{Perdew92prb}.
All calculations are spin-paired with the exception indicated.
For isolated systems, the size of the simulation box had at 
least \unit[4]{\text{\AA}} vacuum around each atom.
Molecules were relaxed until the maximum force dropped 
below \unit[0.05]{eV\text{\AA}$^{-1}$}.

The wave-functions of core atomic states are approximately 
independent of the chemical environment and can be treated 
in the frozen core approximation.
We freeze the $1s$ electrons for C, N, O, and F,
and additionally the $2s, 2p$ electrons for Al, Si, P and S.
The atomic Kohn-Sham states are obtained within a non-spin polarized
spherical symmetric approximation.
A similar approach is adopted to describe the core hole
by lowering the occupation of the relevant state in the
atomic calculation by unity. The resulting Kohn-Sham orbitals
are then used to construct the frozen core
\cite{Enkovaara10jpc,Ljungberg11jesrp}.

{\em Molecules.}
%
We have calculated the $1s$ core hole binding 
energies of the  elements C, N, O, and F and the 
$2p$ core hole energies of Al, Si, P and S for a total of 
63 molecules and 133 chemically distinct core hole binding energies.
Fig.~\ref{fig:molecules} shows that there is an excellent linear correlation between calculated and experimental~\cite{Jolly1984} data.
The main difference between experiment, $E_B^{\rm exp}$,
and simulation, $E_B^{\rm calc}$,
is a constant offset $\delta=E_B^{\rm calc}-E_B^{\rm exp}$ 
that depends on element and core electron level
as demonstrated by the  
straight lines in Fig.~\ref{fig:molecules}.
Binding energies typically span several eV over which the deviation
from a linear correlation analysis is small.

\begin{table}
  \begin{tabular}{l |c |c |c |c |c |c |c |c  }
    functional & C     & N     & O     & F     & Al   & Si    & P     & S\\
    \hline
    PBE     & 6.07 & 6.95  & 7.66  & 8.44  & 0.74 & 0.76  & 0.67  & 0.91 \\
    PBE$^b$ & 6.10 & 7.12  & 7.88  & 8.55  & 0.74 & 0.76  & 0.66  & 0.87 \\
    LDA     & 3.05  & 3.38  & 3.60  & 3.84  & 0.68 & 0.72  & 0.66  & 0.91 \\
    PBE$^c$ & -0.93 & -1.12 & -1.45 & -1.69 & -0.02 & -0.10 & -0.28 & -0.13
  \end{tabular}
  \caption{\label{tab:shifts}
    Empirical offsets $\delta$ in eV between experimental and calculated core hole 
    binding energies as depicted in Fig.~\ref{fig:molecules}. 
    $^b$ PBE calculation with grid spacing $h=0.15$ \AA. 
    $^c$ PBE calculation with corrected spin density (see text).
}
\end{table} 

The values of the offsets $\delta$ 
are given in Fig.~\ref{fig:molecules} and in Tab. \ref{tab:shifts}.
The PBE offsets are always positive (i.e.~$E_B$ is predicted too large),
are largest for the 1s core holes and 
increase from C ($\sim 6\,\text{eV}$) to F ($\sim 8.5\,\text{eV}$) 
as binding energy increases and the core hole becomes more localized.
The offsets are much smaller ($\sim 1\,\text{eV}$) for the 
less localized $2p$ core holes in Al, Si, P and S.
The standard deviations reported in Fig.~\ref{fig:molecules} 
are always $\le 0.1$ eV.
%

The main contribution to $\delta$ comes from neglecting the 
spin dependence of the core hole.
Including the spin dependent core hole density \cite{Ljungberg11jesrp}
yields much 
smaller offsets of $\lesssim 1\,\text{eV}$ (see Tab.~\ref{tab:shifts}).
This computationally more demanding
approach leads to rather small corrections for the $2p$ holes, 
but it overcorrects for the more localized $1s$ holes.
%
Table ~\ref{tab:shifts} also
shows that $\delta$ is strongly dependent on the functional.
LDA gives much lower values than PBE.
The qualitative trend of increasing values from C to F and lower
offsets for $2p$ holes prevails.

\begin{figure*}[htb]
  \includegraphics[width=\linewidth]{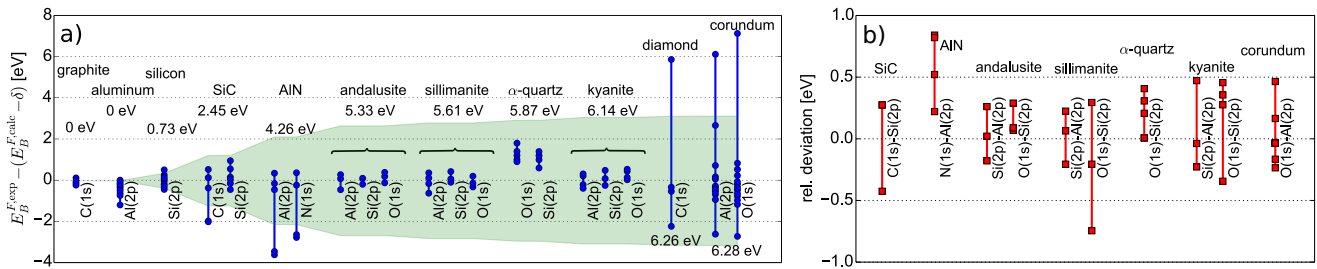}
  \caption{
    \label{fig:Solids}
    a) Difference between experimental $E_B^{F, {\rm exp}}$ and 
    calculated $E_B^{F, {\rm calc}}-\delta$ for 
    various solids. The numbers quoted and the shaded area give the 
    PBE band gaps of the respective material.
    b) Differences of relative shifts between pairs of elements for a certain compound.
    Multiple points indicate individual experimental results.
  }
\end{figure*}

Numerical settings influence the values of the offsets
\cite{Walter15prb}, where
the grid spacing $h$ is found to have 
the largest influence as can be seen in Table \ref{tab:shifts}.
The remaining differences to the experiment 
might have several reasons including dynamic 
screening effects~\cite{Kotsis06pccp,Susi14bjn},
spin-orbit contributions and 
inaccuracies of the density 
functional applied.

The important observation is, that the offset $\delta$
does depend on just atom type and orbital and is --
to a good approximation -- independent of molecule and chemical environment.
This enables a correction of energies obtained from 
facile spin 
unpolarized calculations for an accurate prediction of experimental 
core hole binding energies.
The standard deviation of the approach is
smaller than 0.2 eV for all elements considered, an accuracy that
is sufficient in comparison to the variation over several eV
due to differences in the chemical environment. 
 
{\em Bulk solids.}
The $\delta$-values parameterized from molecular spectra can now be used to
predict CEBEs of bulk solids.
We used experimental lattice constants and atomic positions for 
all solids studied. 
The complete reference list of the extensive experimental data
used in this study is found in Supplemental Material \cite{supplemental}.
An extra electron was added to the valence band while 
keeping the core hole empty and
$E_B^F$ was evaluated according to eq. (\ref{eq:echF0}).
Fig.~\ref{fig:Solids}a) shows the resulting differences between experimental
and calculated CEBEs for various solids.
Including the offset, we find 
our calculated {\it absolute} CEBE to be in excellent
agreement with experiment for the semi metal
bulk graphite and the metal 
aluminum,
where the Fermi level 
serves as good reference.

This is not the case anymore for non-metals
as is most obvious
for materials with the largest gaps, such as diamond and corrundum, 
where experimental values spread over many eV.
Despite of this problem, we find an astonishing agreement
even in absolute CEBES for many systems by setting 
$\varepsilon_F=(\varepsilon_H+\varepsilon_L)/2$ 
using PBE values for $\varepsilon_H$ and $\varepsilon_L$.
This approach is satisfactory compared to the experimental spread for
crystalline silicon and 
diamond as well as crystalline compounds of 
Al, Si, C and O, in particular nitrides (AlN), 
oxides (corrundum), 
carbides (SiC) and aluminosilicates (andalusite, 
kyanite and sillimanite which are 
polymorphs of Al$_2$SiO$_5$).
The only exception is $\alpha$-quartz, where the 
calculated values are too small by roughly 1 eV as compared to experiment.

Energy reference problems cancel out when CEBE
differences within a single compound
are considered.
In this case the experimental energy spread is drastically
reduced when values from the same experiment are compared 
(Fig.~\ref{fig:Solids}b)).
Also the agreement between prediction and experiment is excellent
for all solids considered proving that
the prediction of relative shifts is possible even for systems
with large gaps.
Note that the empirical energy offsets $\delta$ 
do not cancel since they vary between 
different elements and core levels and need to be 
considered even for the calculation of relative shifts.

{\em Summary \& Discussion.}
We have presented a convenient computational method for the prediction of
absolute core electron binding energies (CEBEs) from density-functional theory 
(DFT) calculations within the projector augmented wave 
(PAW) formulation.
A large set of experimental molecular data have been used to show that
spin paired gas phase calculations predict CEBES in excellent accuracy
up to a constant offset that mainly depends 
on chemical element and core level.
These offsets allow an \emph{a-posteriori} correction of DFT results 
and therefore a prediction of core
electron binding energies also for solids.
The main issue in absolute CEBEs is the definition of the reference energy,
a problem also present in experiment.
This mainly affects systems with a band
gap, where the common Fermi energy reference is not defined.
Chemical shifts, that do not require the definition of
an energy reference can still be calculated with good accuracy.

It would be desirable to have $\delta$ values for all elements
of the chosen functional. Such a database could serve for the
assignment of experimental XPS peaks via direct comparison to simulations. 
Gas-phase XPS spectra can
be used for parametrisation as shown above. 
For elements where these are not available, 
an alternative route is the calibration from metallic alloys
where accurate relative shifts are available.
Finally also more accurate, but computationally more demanding
quantum chemical calculations could be used to obtain $\delta$ without
reference to experiment.

{\em Acknowledgments.}
Computational resources of FZ-Jülich are thankfully acknowledged. 
This work is partially based on research funded by the DFG
(grants PA 2023/2 and MO 879/17).


%

\end{document}